\documentclass{article}%
\usepackage{amsmath}
\usepackage{amssymb}
\usepackage{amsfonts}
\usepackage{graphicx}%
\providecommand{\U}[1]{\protect\rule{.1in}{.1in}}
\newtheorem{theorem}{Theorem}
\newtheorem{acknowledgement}[theorem]{Acknowledgement}

\begin{document}

\author{Richard Healey\\University of Arizona}
\title{Quantum Theory and the Limits of Objectivity}
\date{September 2018}
\maketitle

\begin{abstract}
Three recent arguments seek to show that the universal applicability of
unitary quantum theory is inconsistent with the assumption that a
well-conducted measurement always has a definite physical outcome. In this
paper I restate and analyze these arguments. The import of the first two is
diminished by their dependence on assumptions about the outcomes of
counterfactual measurements. But the third argument establishes its intended
conclusion. Even if every well-conducted quantum measurement we ever make will
have a definite physical outcome, this argument should make us reconsider the
objectivity of that outcome.

\end{abstract}

\section{Introduction}

Quantum theory is taken to be fundamental to contemporary physics in large
part because countless measurements have yielded outcomes that conform to its
predictions. Experimenters take great care to ensure that each quantum
measurement has an outcome that is not just a subjective impression but an
objective, physical event. However, in the continuing controversy in quantum
foundations QBists (\cite{Fuchs}, \cite{FuchsMerminSchack}) and others
(\cite{Brukner}, \cite{Brukner 2}, \cite{Rovelli}) have come to question and
even deny the principle that a well-conducted quantum measurement has a
definite, objective, physical outcome. This principle should not be abandoned
lightly: objective data provide the platform on which scientific knowledge
rests.\footnote{Even if no item of data is so certain as to be immune from
rejection in the light of further scientific investigation. Recall Popper's
(\cite{Popper}, p. 94) famous metaphor:
\par
\textquotedblleft Science does not rest upon solid bedrock. The bold structure
of its theories rises, as it were, above a swamp. It is like a building
erected on piles. The piles are driven down from above into the swamp, but not
down to any natural or `given' base; and if we stop driving the piles deeper,
it is not because we have reached firm ground. We simply stop when we are
satisfied that the piles are firm enough to carry the structure, at least for
the time being.\textquotedblright} We should demand a water-tight argument
before giving it up.

In this paper I analyze three recent arguments that quantum theory,
consistently applied, entails that not every quantum measurement can have a
definite, objective, physical outcome. I say `can have', not `has', because
each argument requires a \textit{Gedankenexperiment} far more extreme even
than that of Schr\"{o}dinger's cat. The first two arguments' dependence on
questionable implicit assumptions severely limits their significance. But I
think the third argument at least succeeds in deflating a certain ideal of
objectivity in the quantum domain. I assume throughout that an outcome of a
quantum measurement is definite only if it is unique---an assumption rejected
by Everettians such as Deutsch \cite{Deutsch} and Wallace \cite{Wallace}.
Assuming the objectivity of a physical outcome, an Everettian may take an
argument like these considered here as offered in support of that outcome's
non-uniqueness, as suggested by the title of \cite{Renner}.

\section{Brukner's Argument}

Brukner's argument (\cite{Brukner}, \cite{Brukner 2}) applies Bell's theorem
\cite{Bell} to an\ extension of Wigner's \cite{Wigner} friend scenario. My
restatement of the most recent version \cite{Brukner 2} of his
argument\ renames Brukner's characters and introduces clarifying notation.

Before describing his own \textit{Gedankenexperiment}, Brukner considers
Deutsch's \cite{Deutsch} twist on Wigner's original friend scenario. So
consider first a scenario in which Zeus\footnote{Brukner calls this character
Wigner, but I have reserved that name for another character with analogous
powers.} is contemplating possible measurements on Xena's otherwise physically
isolated lab $X$,\ inside which Xena has measured the $z$-spin of a single
spin-1/2 particle $1$\ prepared in the superposed state in the $z$-spin basis
\begin{equation}
\left\vert x\right\rangle _{1}=1/\sqrt{2}(\left\vert \uparrow\right\rangle
_{1}+\left\vert \downarrow\right\rangle _{1})
\end{equation}
Assuming the universal applicability of unitary quantum mechanics, Zeus
assigns to the combined system $1X$ after Xena's measurement the entangled
state%
\begin{equation}
\left\vert \Phi\right\rangle _{1X}=1/\sqrt{2}(\left\vert \uparrow\right\rangle
_{1}\left\vert \text{\textquotedblleft up\textquotedblright}\right\rangle
_{X}+\left\vert \downarrow\right\rangle _{1}\left\vert \text{\textquotedblleft
down\textquotedblright}\right\rangle _{X}),
\end{equation}
where $\left\vert \text{\textquotedblleft up\textquotedblright}\right\rangle
_{X}$ (for example) represents a state in which if Zeus were to observe the
contents of Xena's lab he would certainly (with probability $1$) find her
reporting the outcome of her measurement of $z$-spin on particle $1$ as
$+\hbar/2$ and that her recording device had indeed recorded that value.

Zeus can try to verify his assignment of state $\left\vert \Phi\right\rangle
_{1X}$\ by performing a measurement of a dynamical variable $A_{x}%
$\ represented by the operator $\hat{A}_{x}$ on $H_{1}\otimes H_{X}$%
\begin{equation}
\hat{A}_{x}=\left\vert \uparrow\right\rangle _{1}\left\vert
\text{\textquotedblleft up\textquotedblright}\right\rangle _{X}\left\langle
\downarrow\right\vert _{1}\left\langle \text{\textquotedblleft
down\textquotedblright}\right\vert _{X}+\left\vert \downarrow\right\rangle
_{1}\left\vert \text{\textquotedblleft down\textquotedblright}\right\rangle
_{X}\left\langle \uparrow\right\vert _{1}\left\langle \text{\textquotedblleft
up\textquotedblright}\right\vert _{X}\text{.}%
\end{equation}
This measurement will (with probability $1$) yield the outcome $+1$ while
leaving the state $\left\vert \Phi\right\rangle _{1X}$ undisturbed. After
this\ successful verification, Zeus's apparatus and memory establish the truth
of statement $A_{x}^{+}$: \textquotedblleft Zeus's outcome is $A_{x}=+1
$\textquotedblright\ and the falsity of $A_{x}^{-}$: \textquotedblleft Zeus's
outcome is $A_{x}=-1$.\textquotedblright\ But despite his entangled state
assignment, Zeus may have some reason to believe that Xena has indeed observed
a definite outcome of her measurement of $z$-spin.

As Deutsch \cite{Deutsch}\ pointed out, no violation of unitary quantum theory
is involved if Xena passes a message out of her lab to Zeus reporting that she
has seen a definite outcome, as long as this contains no information about
what that outcome was. Zeus may try to see for himself whether Xena has seen a
definite outcome by performing his own measurement on her lab and its
contents, of a dynamical variable $A_{z}$\ represented by the operator
$\hat{A}_{z}$ on $H_{1}\otimes H_{X}$%
\begin{equation}
\hat{A}_{z}=\left\vert \uparrow\right\rangle _{1}\left\vert
\text{\textquotedblleft up\textquotedblright}\right\rangle _{X}\left\langle
\uparrow\right\vert _{1}\left\langle \text{\textquotedblleft
up\textquotedblright}\right\vert _{X}-\left\vert \downarrow\right\rangle
_{1}\left\vert \text{\textquotedblleft down\textquotedblright}\right\rangle
_{X}\left\langle \downarrow\right\vert _{1}\left\langle
\text{\textquotedblleft down\textquotedblright}\right\vert _{X}\text{.}%
\end{equation}
If Zeus's outcome is $A_{z}=+1$ he may judge this to verify the statement
$A_{z}^{+}$: \textquotedblleft Xena's outcome is $z^{+}$,\textquotedblright%
\ and falsify $A_{z}^{-}$: \textquotedblleft Xena's outcome is $z^{-}%
$\textquotedblright, while outcome $A_{z}=-1$ reverses these judgments. These
judgments are not warranted by the (false) assumption that an ideal quantum
measurement just faithfully reveals the pre-existing value of the measured
variable. Instead, their warrant rests on the assumption that Xena's outcome
is accessible to other observers by consulting her records. Failure of such
intersubjectivity would undermine Xena's outcome's claim to objectivity, at
least in this epistemic sense.

Since the measurement of $A_{x}$\ leaves the state $\left\vert \Phi
\right\rangle _{1X}$ unchanged, Zeus may first perform that measurement to
establish the truth of $A_{x}^{+}$, then measure $A_{z}$ to verify the truth
of $A_{z}^{+}$ (or, alternatively, of $A_{z}^{-}$). So in this preliminary
scenario Zeus has some reason to believe that not only his own measurements
but also Xena's measurement had a definite, physical outcome. Moreover, if he
measures only $A_{x}$ he can then pass a message with its outcome to Xena,
also without disturbing the state $\left\vert \Phi\right\rangle _{1X}$ :
so\ Xena, too, will have reason to believe that both $A_{x}^{+}$ and
$A_{z}^{+}$ (or $A_{z}^{-}$) are true and that Zeus's measurement of $A_{x}%
$\ as well as her own measurement of $1$'s $z$-spin had a definite, physical outcome.

Now consider the statements $c(A_{x}^{+})$: \textquotedblleft$A_{x}^{+}$ would
be true if Zeus were to measure $A_{x}$\textquotedblright, $c(A_{x}^{-})$:
\textquotedblleft$A_{x}^{-}$ would be true if Zeus were to measure $A_{x}%
$\textquotedblright; $c(A_{z}^{+})$: \textquotedblleft Zeus's outcome would be
$A_{z}=+1$\ if he were to measure $A_{z}$\textquotedblright, $c(A_{z}^{-})$:
\textquotedblleft Zeus's outcome would be $A_{z}=-1$\ if he were to measure
$A_{z}$\textquotedblright. Zeus has reason to believe $c(A_{x}^{+})$ is true
and $c(A_{x}^{-})$ is false whether or not he measures $A_{x}$, since
$\left\vert \Phi\right\rangle _{1X}$ predicts the truth of $A_{x}^{+}$ (with
probability $1$). Whether or not he measures $A_{z}$, Zeus has reason to
believe that one of $c(A_{z}^{+})$, $c(A_{z}^{-})$ is true while the other is
false in state $\left\vert \Phi\right\rangle _{1X}$. Assuming measurements
have definite, objective outcomes, he should take his conditional outcome
simply to reflect Xena's actual outcome: Xena got $z^{+}$ if and only if Zeus
would get $+1$, while Xena got $z^{-}$ if and only if Zeus would get $-1$.
Provided that Xena's measurement had a definite actual outcome it follows that
exactly one of $c(A_{z}^{+})$ or $c(A_{z}^{-})$ is true.

After analyzing this preliminary scenario, Brukner \cite{Brukner
2}\ introduces his own, more complex, \textit{Gedankenexperiment}. Each of
Xena and Yvonne is located in a separate laboratory. These laboratories are
initially completely physically isolated, and this isolation is preserved
except for the processes specified below. An entangled pair of spin-$1/2$
particles is prepared, with particle $1$ in Xena's lab $X$ and particle $2$ in
Yvonne's lab $Y$. In \cite{Brukner 2}\ the initial state assigned to $12$ (in
the $z$-spin basis) is
\begin{align}
\left\vert \psi\right\rangle _{12}  &  =-\sin\theta/2\left\vert \phi
^{+}\right\rangle _{12}+\cos\theta/2\left\vert \psi^{-}\right\rangle
_{12}\text{, where}\\
\left\vert \phi^{+}\right\rangle _{12}  &  =1/\sqrt{2}(\left\vert
\uparrow\right\rangle _{1}\left\vert \uparrow\right\rangle _{2}+\left\vert
\downarrow\right\rangle _{1}\left\vert \downarrow\right\rangle _{2}%
)\nonumber\\
\left\vert \psi^{-}\right\rangle _{12}  &  =1/\sqrt{2}(\left\vert
\uparrow\right\rangle _{1}\left\vert \downarrow\right\rangle _{2}-\left\vert
\downarrow\right\rangle _{1}\left\vert \uparrow\right\rangle _{2}).\nonumber
\end{align}
Then Xena measures the $z$-spin of particle $1$\ in her lab, while Yvonne
measures the $z$-spin of particle $2$\ in her lab. Assume that the measurement
in each laboratory has a definite, physical outcome, registered by a particle
detector, recorded in a computer (or on paper) and experienced by Xena or
Yvonne respectively.

Each of Zeus and Wigner is also located in a separate laboratory. Xena's
laboratory is located wholly within Zeus's, while Yvonne's is located wholly
within\ Wigner's. But to this point each laboratory has remained completely
physically isolated insofar as there has been no direct physical interaction
between any of these four laboratories.

Assuming (no-collapse) quantum theory is universally applicable, there is a
correct quantum state for Zeus and Wigner to assign to the joint physical
system consisting of the entire contents of both Xena's and Yvonne's
laboratories and this state evolved unitarily throughout the interactions
involved in each of their spin-component measurements. (Note that in assigning
this state, Zeus and Wigner are here treating Xena and Yvonne themselves as
quantum (sub)systems.) Assuming for simplicity that the spin-component
measurements were non-disturbing, we may write this joint state after Xena's
and Yvonne's measurements as%
\begin{align}
\left\vert \Psi\right\rangle _{12XY}  &  =-\sin\theta/2\left\vert \Phi
^{+}\right\rangle +\cos\theta/2\left\vert \Psi^{-}\right\rangle \text{,
where}\label{Big entangled state}\\
\left\vert \Phi^{+}\right\rangle  &  =1/\sqrt{2}\left(  \left\vert
A_{up}\right\rangle \left\vert B_{up}\right\rangle +\left\vert A_{down}%
\right\rangle \left\vert B_{down}\right\rangle \right) \nonumber\\
\left\vert \Psi^{-}\right\rangle  &  =1/\sqrt{2}\left(  \left\vert
A_{up}\right\rangle \left\vert B_{down}\right\rangle -\left\vert
A_{down}\right\rangle \left\vert B_{up}\right\rangle \right)  .\nonumber
\end{align}
Here $X$ represents the entire contents of Xena's lab (including Xena) and $Y
$ represents the entire contents of Yvonne's lab (including Yvonne), except
the measured particles $1,2$. $\left\vert A_{up}\right\rangle $, $\left\vert
A_{down}\right\rangle $ are eigenstates of $\hat{A}_{z}$.

We may define an analogous pair of self-adjoint operators on $H_{2}\otimes
H_{Y}$\ as follows:%
\begin{align*}
\hat{B}_{z}  &  =\left\vert \uparrow\right\rangle _{2}\left\vert
\text{\textquotedblleft up\textquotedblright}\right\rangle _{Y}\left\langle
\uparrow\right\vert _{2}\left\langle \text{\textquotedblleft
up\textquotedblright}\right\vert _{Y}-\left\vert \downarrow\right\rangle
_{2}\left\vert \text{\textquotedblleft down\textquotedblright}\right\rangle
_{Y}\left\langle \downarrow\right\vert _{2}\left\langle
\text{\textquotedblleft down\textquotedblright}\right\vert _{Y}\\
\hat{B}_{x}  &  =\left\vert \uparrow\right\rangle _{2}\left\vert
\text{\textquotedblleft up\textquotedblright}\right\rangle _{Y}\left\langle
\downarrow\right\vert _{2}\left\langle \text{\textquotedblleft
down\textquotedblright}\right\vert _{Y}+\left\vert \downarrow\right\rangle
_{2}\left\vert \text{\textquotedblleft down\textquotedblright}\right\rangle
_{Y}\left\langle \uparrow\right\vert _{2}\left\langle \text{\textquotedblleft
up\textquotedblright}\right\vert _{Y}%
\end{align*}
where magnitude $B_{z}$ on $2Y$ uniquely corresponds to $\hat{B}_{z}$ and
$B_{x}$ to $\hat{B}_{x}$. The state (\ref{Big entangled state}) predicts that
the statistics of the (assumed, definite) outcomes of Zeus's and Wigner's
measurements will violate the associated Clauser-Horne-Shimony-Holt inequality%
\begin{equation}
S=\left\langle A_{z}B_{z}\right\rangle +\left\langle A_{z}B_{x}\right\rangle
+\left\langle A_{x}B_{z}\right\rangle -\left\langle A_{x}B_{x}\right\rangle
\leq2 \tag{CHSH}\label{CHSH}%
\end{equation}
in which $\left\langle A_{x}B_{z}\right\rangle $, for example, is the
correlation function of a probability distribution for the outcomes of
measurements of magnitudes $A_{x},B_{z}$. The inequality \ref{CHSH} is
violated, for example, by state $\left\vert \Psi\right\rangle _{12XY}$ for
which $S=2\sqrt{2}$ if $\theta=\pi/4$.

Suppose that the sequence of measurements by Xena, Yvonne, Zeus and Wigner is
repeated in many trials, with Zeus's measurement of $A_{x}$ or $A_{z}$\ and
Wigner's measurement of $B_{x}$ or $B_{z}$ varied randomly and independently
from trial to trial.\ Violation of (\ref{CHSH}) by statistics collected in a
large number of such trials is perfectly consistent with the assumption

\begin{quote}
\textit{Definite Outcomes}: In every such trial each of Xena's, Yvonne's,
Zeus's and Wigner's measurements has a definite, physical outcome.
\end{quote}

The assumption of \textit{Definite Outcomes} does not even make it unlikely
that a large number of Zeus's and Wigner's outcomes in repeated trials will
display correlations in violation of (\ref{CHSH}). Indeed the Born rule
\textit{predicts} that the outcomes of Zeus's and Wigner's measurements will
violate \ref{CHSH}: if $\theta=\pi/4$ then $S=2\sqrt{2}$. Why might one think otherwise?

Brukner \cite{Brukner 2} takes his argument to disprove the following postulate

\begin{quote}
\textbf{Postulate} (\textquotedblleft Observer-independent
facts\textquotedblright) \textit{The truth-values of the propositions }$A_{i}$
\textit{of all observers form a Boolean algebra} $\mathcal{A}$.
\textit{Moreover, the algebra is equipped with a (countably additive) positive
measure }$p(A)\geqq0$ \textit{for all statements }$A\in\mathcal{A} $,
\textit{which is the probability for the statements to be true.}
\end{quote}

To evaluate the bearing of his argument on the assumption of \textit{Definite
Outcomes}\ one must specify propositions purporting to describe such outcomes.
Brukner's discussion of the preliminary scenario suggests these include
$A_{z}^{+}$, $A_{z}^{-}$, $A_{x}^{+}$, and $A_{x}^{-}$. The symmetry of the
\textit{Gedankenexperiment} further suggests they include propositions
$B_{z}^{+}$, $B_{z}^{-}$, $B_{x}^{+}$ and $B_{x}^{-}$, each of which states
the outcome of an analogous measurement by Yvonne or by Wigner. Is there any
reason to believe that application of Brukner's \textbf{Postulate} to the
propositions $\mathcal{B}=\{A_{z}^{+},A_{z}^{-},B_{z}^{+},B_{z}^{-},A_{x}%
^{+},A_{x}^{-},B_{x}^{+},B_{x}^{-}\}$ yields the promised no-go therem?

In no repetition are both $A_{x}$ and $A_{z}$ measured---the experimental
arrangements are mutually exclusive, as are those for $B_{x}$ and $B_{z}$. If
$A_{x}$ is not measured, then neither $A_{x}^{+}$ nor $A_{x}^{-}$ describes an
actual outcome: and if $B_{x}$ is not measured, then neither $B_{x}^{+}$ nor
$B_{x}^{-}$ describes an actual outcome. So unless $A_{x}$, $B_{x}$ are
measured in a repetition, the propositions of all observers that describe the
actual definite outcomes assumed by \textit{Definite Outcomes} is not the
whole of $\mathcal{B}$\ but merely a compatible subset $\mathcal{B}^{\ast}$
forming a Boolean algebra which may readily be equipped with a (countably
additive) positive measure: just use the Born probabilities from state
(\ref{Big entangled state}) and extend this to each proposition describing the
outcome of an actual measurement by Xena or by Yvonne by equating its outcome
to that of the corresponding measurement by Zeus or by Wigner (so, for
example, $A_{z}^{+}$ is true if and only if the outcome of Zeus's measurement
of $A_{z}$ is $A_{z}=+1$, and both propositions have the same probability).

If $A_{x}$, $B_{x}$ are measured in a repetition, the propositions of all
observers describing the actual definite outcomes assumed by \textit{Definite
Outcomes} is the whole of $\mathcal{B}$. But the propositions $\{A_{x}%
^{+},A_{x}^{-},B_{x}^{+},B_{x}^{-}\}$\ form a Boolean algebra whose structure
is respected by the obvious truth-assignments, and the Born probabilities from
(\ref{Big entangled state}) define a probability measure on this algebra. In
the absence of any further constraints it is easy to extend this
truth-assignment and probability measure to the full algebra $\mathcal{B}$.

So the assumed actual outcomes in each trial can certainly be described by
propositions $A_{i}$ of all observers forming a Boolean algebra $\mathcal{A}$.
Moreover, this algebra may be equipped with a (countably additive) positive
measure $p(A)\geqq0$ for all statements $A\in A$, which may be taken as the
probability for the statements to be true in that trial. A no-go therorem is
not derivable through the application of \textit{Definite Outcomes} to
propositions $A_{i}$ of all observers that describe their actual outcomes in
any, or all, repetitions of Brukner's \textit{Gedankenexperiment}.

What happens if instead the \textquotedblleft propositions of all
observers\textquotedblright\ concern not their actual but their hypothetical
outcomes? Consider the set\newline$c(\mathcal{B)}=\{c(A_{z}^{+}),c(A_{z}%
^{-}),c(B_{z}^{+}),c(B_{z}^{-}),c(A_{x}^{+}),c(A_{x}^{-}),c(B_{x}^{+}%
),c(B_{x}^{-})\}$ of subjunctive conditionals describing the outcomes of
hypothetical measurements. Assume that if such a measurement is actually made
in a trial then the corresponding conditional has the same truth-value as its
consequent (so, for example, if $A_{x}$ is measured with outcome $A_{x}=+1$
then $c(A_{x}^{+})$ is true as well as $A_{x}^{+}$). Unlike the simpler
scenario discussed earlier, when $A_{x}^{+},A_{x}^{-}$ are replaced by the
corresponding subjunctive statements $c(A_{x}^{+}),c(A_{x}^{-})$:
\textquotedblleft If $A_{x}$ were measured then the definite outcome would be
$+1$ ($-1$)\textquotedblright,\ in state (\ref{Big entangled state})\ there is
no reason to suppose that either of these statements even \textit{has }a
truth-value if Zeus does not measure $A_{x}$. Nor should $c(B_{x}^{+}%
),c(B_{x}^{-})$ be expected to have truth-values when Wigner does not measure
$B_{x}$.

Unless $A_{x}$,$B_{x}$ are both measured in a trial, replacement of
propositions about actual definite outcomes of a measurement by such
conditionals fails to generate a Boolean algebra of propositions of all
observers whose truth-value assignment respects that algebra. But quantum
theory predicts violation of the inequality \ref{CHSH} only for the outcomes
of actual measurements. Because of the physical incompatibility of Zeus's
joint measurement of $A_{x}$ and $A_{z}$\ and of Wigner's measurement of
$B_{x}$ and $B_{z}$, these predictions must concern four distinct kinds of
trials, which is what necessitated variation of measurements by Zeus and by
Wigner from trial to trial. \textit{Definite Outcomes} implies that the set
$c(\mathcal{B)}$ forms a Boolean algebra whose structure is respected by a
joint truth-assignment and is equipped with a (countably additive) positive
measure \textit{at most }in the case of a repetition in which Zeus measures
$A_{x}$ and Wigner measures $B_{x}$. So the violation of that inequality in
state (\ref{Big entangled state}) does not refute \textit{Definite Outcomes.
As it stands, }Brukner's argument (\cite{Brukner}, \cite{Brukner 2}) provides
no good reason to doubt that every quantum measurement has a definite,
objective, physical outcome.

In correspondence, Brukner has proposed a slight modification that avoids this
objection and promises to strengthen the argument. In the modified scenario,
Zeus measures $A_{x}$ and Wigner measures $B_{x}$ in \textit{every} trial. As
in the simpler Wigner's friend scenario, Zeus may appeal to the epistemic
objectivity of Xena's outcome to infer that $c(A_{z}^{+})$ has the same
truth-value as $A_{z}^{+}$, and $c(A_{z}^{-})$ has the same truth-value as
$A_{z}^{-}$.\footnote{Though this inference is now questionable, since in this
context the antecedent "Zeus measures $A_{z}$" of the counterfactuals
$c(A_{z}^{+}),c(A_{z}^{-})$\ is not merely false but incompatible with Zeus's
actual measurement of $A_{x}$.} Since \textit{Definite Outcomes} implies
that\ one of $A_{z}^{+},A_{z}^{-}$ is true and the other false, it then
follows that in every repetition one of $c(A_{z}^{+}),c(A_{z}^{-})$ is true
and the other false, even though Zeus actually measures $A_{x}$ and not
$A_{z}$ in that repetition. Similarly, in every repetition one of $c(B_{z}%
^{+}),c(B_{z}^{-})$ is true and the other is false. So in each repetition the
set of propositions $\{c(A_{z}^{+}),c(A_{z}^{-}),c(B_{z}^{+}),c(B_{z}^{-})\}$
always forms a Boolean algebra whose truth-value assignment and probability
distribution follow from those assigned to the assumed actual outcomes of
Xena's and Yvonne's measurements. This will be true in every trial of this
modified scenario.

\textit{Definite Outcomes} now implies that every proposition in the full
algebra $c(\mathcal{B})$ has a truth-value and these truth-values respect the
algebra's structure. Moreover, any (countably additive) positive
measure\textit{ }$p(A)\geqq0$ for all statements in\textit{ }$c(\mathcal{B})$
must be constrained by a transformed inequality obtained from \ref{CHSH} by
replacing each reference to an actual outcome by a reference to the
corresponding hypothetical outcome (though for $A_{x}$,$B_{x}$ the
hypothetical outcome \textit{is} the actual outcome). If quantum theory were
to predict violation of this transformed inequality then it would imply that
\textit{Definite Outcomes} is false.

But quantum theory predicts probabilities only for the outcomes of actual
measurements, and neither $A_{z}$ nor $B_{z}$ is actually measured in this
modified scenario. Only $A_{x}$,\ $B_{x}$ and the $z$-spins of $\mathbf{1}%
$,$\mathbf{2}$ are measured in each trial, and quantum theory makes no
predictions of the joint probability distribution for Zeus's and Yvonne's
pairs of measurement outcomes, or that for Wigner's and Xena's pairs of
measurement outcomes. This is to be expected, since even if \textit{Definite
Outcomes} is true, these outcome pairs are not epistemically accessible by any
observer (including the four agents named in this scenario), so their
statistics are of no scientific interest.

\section{Frauchiger and Renner's Argument}

Here is a simplified restatement of the argument of Frauchiger and Renner
(\cite{Renner}, \cite{Renner 2}). The appendix compares its strategy to that
of the arguments on which it is based and supplies a translation to the
notation of \cite{Renner 2}.

Four physical observers are each located in their own separate laboratories.
Every laboratory is initially completed physically isolated, and this
isolation is preserved except for the processes specified below. In one
laboratory observer Xena has prepared a "biased quantum coin" $c$ in state%
\begin{equation}
\left\vert ready\right\rangle _{c}=\frac{1}{\sqrt{3}}\left\vert
heads\right\rangle _{c}+\frac{\sqrt{2}}{\sqrt{3}}\left\vert tails\right\rangle
_{c}.
\end{equation}
At time $t=0$ Xena "flips the coin" by implementing a measurement on $c$ of
observable $f$ with orthonormal eigenstates $\left\vert heads\right\rangle
_{c}$, $\left\vert tails\right\rangle _{c}$ by means of a unitary interaction
with $c$.%
\begin{equation}
\left\vert ready\right\rangle _{c}\left\vert ready\right\rangle _{X^{-}%
}\Longrightarrow\left\vert \psi\right\rangle _{cX^{-}}^{0}=\frac{1}{\sqrt{3}%
}\left\vert heads\right\rangle _{c}^{0}\left\vert heads\right\rangle _{X^{-}%
}+\frac{\sqrt{2}}{\sqrt{3}}\left\vert tails\right\rangle _{c}^{0}\left\vert
tails\right\rangle _{X^{-}}. \label{2}%
\end{equation}
Here and elsewhere I put a numerical superscript $n$ on a state to mark its
unitary evolution up to just after time $t=n$. $X^{-}$ is a system
representing the entire contents of Xena's lab (including Xena herself, but
neither $c$ nor a qubit system $s$ whose state she is about to prepare), while
$X$ is $X^{-}+$ $c$. $\left\vert heads\right\rangle _{X^{-}}$, $\left\vert
tails\right\rangle _{X^{-}}$\ are orthonormal eigenstates of a binary
indicator observable on $X$ whose eigenvalue \ $x=1$ represents Xena's outcome
\textquotedblleft heads\textquotedblright\ and whose eigenvalue \ $x=-1$
represents Xena's outcome \textquotedblleft tails\textquotedblright.%
\begin{align}
\hat{x}\left\vert heads\right\rangle _{X^{-}}  &  =\left\vert
heads\right\rangle _{X^{-}}\\
\hat{x}\left\vert tails\right\rangle _{X^{-}}  &  =-\left\vert
tails\right\rangle _{X^{-}}.\nonumber
\end{align}
Assume Xena's measurement of $f$\ on $c$\ has a unique, physical outcome:
either \textquotedblleft heads\textquotedblright\ or \textquotedblleft
tails\textquotedblright.

At time $t=1$, if the outcome was \textquotedblleft heads\textquotedblright,
Xena prepares the state of a qubit system $s$ in her lab in state $\left\vert
\downarrow\right\rangle _{s}$: if the outcome was \textquotedblleft
tails\textquotedblright, Xena prepares $s$ in state $\left\vert \rightarrow
\right\rangle _{s}=1/\sqrt{2}(\left\vert \downarrow\right\rangle
_{s}+\left\vert \uparrow\right\rangle _{s})$. Xena can do this by means of a
unitary interaction between $s$ and $X$, yielding the following state
\begin{align}
\left\vert \psi\right\rangle _{cX^{-}s}^{1}  &  =\frac{1}{\sqrt{3}}\left\vert
heads\right\rangle _{c}^{1}\left\vert heads\right\rangle _{X^{-}}%
^{1}\left\vert \downarrow\right\rangle _{s}+\frac{\sqrt{2}}{\sqrt{3}%
}\left\vert tails\right\rangle _{c}^{1}\left\vert tails\right\rangle _{X^{-}%
}^{1}\left\vert \rightarrow\right\rangle _{s}\\
&  =\frac{1}{\sqrt{3}}\left(  \left\vert heads\right\rangle _{c}^{1}\left\vert
heads\right\rangle _{X^{-}}^{1}\left\vert \downarrow\right\rangle
_{s}+\left\vert tails\right\rangle _{c}^{1}\left\vert tails\right\rangle
_{X^{-}}^{1}\left\vert \downarrow\right\rangle _{s}+\left\vert
tails\right\rangle _{c}^{1}\left\vert tails\right\rangle _{X^{-}}%
^{1}\left\vert \uparrow\right\rangle _{s}\right)  .\text{ \ \ \ \ \ \ \ \ \ }%
\end{align}
Xena then transfers system $s$ out of her lab and into Yvonne's lab, keeping
$c$ in her own lab.

Let $Y^{-}$ be a system consisting of the entire contents of Yvonne's lab
(including Yvonne but not the system $s$ transferred to her by Xena), while
$Y$ is $Y^{-}+$ $s$. At time $t=2$ Yvonne measures observable $S_{z}$ on $s$
with orthonormal eigenstates $\left\vert \downarrow\right\rangle _{s}^{2}$,
$\left\vert \uparrow\right\rangle _{s}^{2}$ by means of another unitary
interaction within her lab, yielding state%
\begin{equation}
\left\vert \psi\right\rangle _{cX^{-}sY^{-}}^{2}=\frac{1}{\sqrt{3}}\left(
\begin{array}
[c]{c}%
\left\vert heads\right\rangle _{c}^{2}\left\vert heads\right\rangle _{X^{-}%
}^{2}\left\vert \downarrow\right\rangle _{s}^{2}\left\vert -1/2\right\rangle
_{Y^{-}}+\\
\left\vert tails\right\rangle _{c}^{2}\left\vert tails\right\rangle _{X^{-}%
}^{2}\left\vert \downarrow\right\rangle _{s}^{2}\left\vert -1/2\right\rangle
_{Y^{-}}\text{ \ \ }+\\
\left\vert tails\right\rangle _{c}^{2}\left\vert tails\right\rangle _{X^{-}%
}^{2}\left\vert \uparrow\right\rangle _{s}^{2}\left\vert +1/2\right\rangle
_{Y^{-}}\text{ \ \ \ \ }%
\end{array}
\right)
\end{equation}
which we can rewrite as%
\begin{equation}
\left\vert \psi\right\rangle _{XY}^{2}=\frac{1}{\sqrt{3}}(\left\vert
heads\right\rangle _{X}\left\vert -1/2\right\rangle _{Y}+\left\vert
tails\right\rangle _{X}\left\vert -1/2\right\rangle _{Y}+\left\vert
tails\right\rangle _{X}\left\vert +1/2\right\rangle _{Y}). \label{13}%
\end{equation}
Let $y$ be a binary indicator observable on $Y$ whose eigenvalue \ $y=1$
represents Yvonne's outcome \textquotedblleft$+1/2$\textquotedblright\ and
whose eigenvalue \ $y=-1$ represents Yvonne's outcome \textquotedblleft$-1/2
$\textquotedblright.%
\begin{align}
\hat{y}\left\vert +1/2\right\rangle _{Y}  &  =\left\vert +1/2\right\rangle
_{Y}\\
\hat{y}\left\vert -1/2\right\rangle _{Y}  &  =-\left\vert -1/2\right\rangle
_{Y}.\nonumber
\end{align}
Assume Yvonne's measurement of $S_{z}$\ on $s$\ has a unique, physical
outcome: either \textquotedblleft$+1/2$\textquotedblright\ or
\textquotedblleft$-1/2$\textquotedblright.

The state (\ref{13}) of $XY$ just after $t=2$ may also be expressed as%
\begin{align}
\left\vert \psi\right\rangle _{XY}^{2}  &  =\frac{1}{\sqrt{3}}(\sqrt
{2}\left\vert fail\right\rangle _{X}\left\vert -1/2\right\rangle
_{Y}+\left\vert tails\right\rangle _{X}\left\vert +1/2\right\rangle
_{Y})\tag{13a}\label{13a}\\
&  =\frac{1}{\sqrt{3}}(\left\vert heads\right\rangle _{X}\left\vert
-1/2\right\rangle _{Y}+\sqrt{2}\left\vert tails\right\rangle _{X}\left\vert
fail\right\rangle _{Y})\tag{13b}\label{13b}\\
&  =\frac{1}{2\sqrt{3}}(3\left\vert fail\right\rangle _{X}\left\vert
fail\right\rangle _{Y}+\left\vert fail\right\rangle _{X}\left\vert
OK\right\rangle _{Y}-\left\vert OK\right\rangle _{X}\left\vert
fail\right\rangle _{Y}+\left\vert OK\right\rangle _{X}\left\vert
OK\right\rangle _{Y})\text{ \ \ \ \ \ \ \ } \tag{13c}\label{13c}%
\end{align}
where the states\ $\left\vert fail\right\rangle _{X}$, $\left\vert
OK\right\rangle _{X}$ are defined by%
\begin{align}
\left\vert OK\right\rangle _{X}  &  =\frac{1}{\sqrt{2}}(\left\vert
heads\right\rangle _{X}-\left\vert tails\right\rangle _{X})\\
\left\vert fail\right\rangle _{X}  &  =\frac{1}{\sqrt{2}}(\left\vert
heads\right\rangle _{X}+\left\vert tails\right\rangle _{X})\nonumber
\end{align}
and the states\ $\left\vert fail\right\rangle _{Y}$, $\left\vert
OK\right\rangle _{Y}$ are defined by
\begin{align}
\left\vert OK\right\rangle _{Y}  &  =\frac{1}{\sqrt{2}}(\left\vert
-1/2\right\rangle _{Y}-\left\vert +1/2\right\rangle _{Y})\\
\left\vert fail\right\rangle _{Y}  &  =\frac{1}{\sqrt{2}}(\left\vert
-1/2\right\rangle _{Y}+\left\vert +1/2\right\rangle _{Y}).\nonumber
\end{align}
At time $t=3$ Zeus measures observable $z$ on $X$ with orthonormal eigenstates
$\left\vert fail\right\rangle _{X}^{3}$, $\left\vert OK\right\rangle _{X}^{3}$
and records a unique, physical outcome: either \textquotedblleft
fail\textquotedblright, or \textquotedblleft OK\textquotedblright. At time
$t=4$ Wigner measures observable $w$ on $Y$ with orthonormal eigenstates
$\left\vert fail\right\rangle _{Y}^{4}$, $\left\vert OK\right\rangle _{Y}^{4}$
and records a unique, physical outcome: either \textquotedblleft
fail\textquotedblright, or \textquotedblleft OK\textquotedblright. Finally, at
$t=5$ Wigner consults Zeus and notes the outcome of his measurement of $z$.

In arriving at the quantum state assignment (\ref{13}) (and its equivalents),
Wigner has correctly applied unitary quantum theory to the specified
interactions. Equation (\ref{13c}) implies that with probability $1/12$
(slightly more than 8\%) the outcomes of Zeus's and Wigner's measurements will
both be \textquotedblleft OK\textquotedblright. We now investigate Wigner's
reasoning about the outcomes of Xena's and Yvonne's measurements in such a case.

\textit{Step 1} At $t=5$ Zeus tells me that the outcome of his measurement of
$z$ on $X$ at $t=3$\ was \textquotedblleft OK\textquotedblright, so I infer
that the unique outcome of his measurement of $z$ on $X$ at $t=3$\ was
\textquotedblleft OK\textquotedblright.

\textit{Step 2} Yvonne measured observable $S_{z}$ on $s$ at time $t=2$. If
her outcome had been \textquotedblleft$-1/2$\textquotedblright\ and not
\textquotedblleft$+1/2$\textquotedblright, then equation (\ref{13a}) implies
(with probability 1) that the unique outcome of Zeus's measurement of $z$ on
$X$\ at $t=3$\ was \textquotedblleft fail\textquotedblright\ and not
\textquotedblleft OK\textquotedblright. But I inferred in step 1 that the
unique outcome of his measurement of $z$ on $X$ at $t=3$\ was
\textquotedblleft OK\textquotedblright. So I now infer (with probability 1)
that the unique outcome of Yvonne's measurement of observable $S_{z}$ on $s$
at time $t=2$ was \textquotedblleft$+1/2$\textquotedblright.

\textit{Step 3} Xena measured observable $f$ on $c$ at $t=0$. If her outcome
had been \textquotedblleft heads\textquotedblright\ and not \textquotedblleft
tails\textquotedblright, then equation (\ref{13}) implies (with probability 1)
that the unique outcome of Yvonne's measurement of $S_{z}$ on $s$ at time
$t=2$ was \textquotedblleft$-1/2$\textquotedblright\ and not \textquotedblleft%
$+1/2$\textquotedblright. But I\ inferred in step 2 that the unique outcome of
Yvonne's measurement of observable $S_{z}$ on $s $ at time $t=2$ was
\textquotedblleft$+1/2$\textquotedblright. So I now infer (with probability 1)
that the unique outcome of Xena's measurement of $f$ on $c$ at $t=0$ was
\textquotedblleft tails\textquotedblright.

\textit{Step 4* }The unique outcome of my measurement of $w$ on $Y$ at $t=4$
was \textquotedblleft OK\textquotedblright. But equation (\ref{13b}) implies
(with probability 1) that if the unique outcome of Xena's measurement of $f$
on $c$ at $t=0$ had been \textquotedblleft tails\textquotedblright, the unique
outcome of my measurement of $w$ on $Y$ at $t=4$ would have been
\textquotedblleft fail\textquotedblright. So I infer that the unique outcome
of Xena's measurement of $f$ on $c$ at $t=0$ was \textquotedblleft
heads\textquotedblright\ and not \textquotedblleft tails\textquotedblright.

Since the conclusion of step 4* contradicts the conclusion of step 3, Wigner's
reasoning has here led to a contradiction. The reasoning depended on several
assumptions, at least one of which must be rejected to restore consistency.
These include the three assumptions:

\textit{Universality} Quantum theory may be applied to all systems, including
macroscopic apparatus, observers and laboratories.

\textit{No collapse} When an observable is measured on a quantum system in a
physically isolated laboratory, the state vector correctly assigned by an
external observer to the combined system+laboratory evolves unitarily throughout.

\textit{Unique outcome} A measurement of an observable has a unique, physical outcome.

Unique outcome corresponds to what Frauchiger and Renner \cite{Renner 2}\ call
(S). The appendix discusses the relation between these three assumptions and
Frauchiger and Renner's assumptions (C), (Q), and (S). But step 4* depends on
an additional assumption that should be questioned and, I argue, rejected:

\textit{Intervention Insensitivity} \ The truth-value of an
outcome-counterfactual is insensitive to the occurrence of a physically
isolated intervening event.

An outcome-counterfactual is a statement of the form $O_{t_{1}}\,\square
\!\rightarrow$ $O_{t_{2}}$ where $O_{t}$ states the outcome of a quantum
measurement at $t$, $t_{1}<t_{2}$, and $A\,\square\!\rightarrow B$ means
\textquotedblleft If $A$ had been the case then $B$ would have been the
case\textquotedblright: An event then intervenes just if it occurs in the
interval $(t_{1},$ $t_{2})$, and it is physically isolated if it occurs in a
laboratory that is then physically isolated from laboratories where $O_{t_{1}%
}$, $O_{t_{2}}$ occur.

To see the problem with step 4* of Wigner's reasoning, focus on the
outcome-counterfactual \textquotedblleft If the unique outcome of Xena's
measurement of $f$ on $c$ at $t=0$ had been \textquotedblleft
tails\textquotedblright, the unique outcome of my measurement of $w$ on $Y$ at
$t=4$ would have been \textquotedblleft fail\textquotedblright%
.\textquotedblright\ Zeus's measurement of $z$ on $X$ at $t=3$ was an
intervening event that occurred in Zeus's laboratory $Z\cup X$\ (taken now to
encompass \ the laboratory $X$ on which he performs his measurement of $z$).
At $t=3$, $Z\cup X$ is still physically isolated from $Y$ and $W$. So the
assumption of \textit{Intervention Insensitivity} would\ license\ step 4* of
Wigner's reasoning.

But \textit{Intervention Insensitivity} actually conflicts with the other
assumptions of the argument. To see why, consider how Wigner should apply
quantum theory to Zeus's measurement of $z$ on $X$\ at $t=3$, in accordance
with \textit{Universality} and \textit{No collapse}. Equation (13a) implies%
\begin{equation}
\left\vert \psi\right\rangle _{XY}^{2}=\frac{1}{\sqrt{3}}\left[  \sqrt
{2}\left\vert fail\right\rangle _{X}\left\vert -1/2\right\rangle _{Y}+\frac
{1}{\sqrt{2}}(\left\vert fail\right\rangle _{X}-\left\vert OK\right\rangle
_{X})\left\vert +1/2\right\rangle _{Y}\right]
\end{equation}
Assume for simplicity that Zeus's measurement on $X$ is non-disturbing. Wigner
knows that Zeus made a non-disturbing measurement of $z$ on $X$ at $t=3$. So
the state he should assign to $XYZ$\ immediately following this measurement is%
\begin{multline}
\left\vert \psi\right\rangle _{XYZ}^{3}=\frac{1}{\sqrt{3}}\left[
\begin{array}
[c]{c}%
\sqrt{2}\left\vert fail\right\rangle _{X}^{3}\left\vert
\text{\textquotedblleft}fail\text{\textquotedblright}\right\rangle
_{Z}\left\vert -1/2\right\rangle _{Y}^{3}\text{ \ \ }%
+\text{\ \ \ \ \ \ \ \ \ \ \ \ \ \ \ \ \ \ \ \ \ \ }\\
+\frac{1}{\sqrt{2}}(\left\vert fail\right\rangle _{X}^{3}\left\vert
\text{\textquotedblleft}fail\text{\textquotedblright}\right\rangle
_{Z}-\left\vert OK\right\rangle _{X}^{3}\left\vert \text{\textquotedblleft%
}OK\text{\textquotedblright}\right\rangle _{Z})\left\vert +1/2\right\rangle
_{Y}^{3}%
\end{array}
\right] \\
=\frac{1}{\sqrt{24}}\left[
\begin{array}
[c]{c}%
\left\vert heads\right\rangle _{X}^{3}\left\{
\begin{array}
[c]{c}%
\left(  3\left\vert fail\right\rangle _{Y}^{3}+\left\vert OK\right\rangle
_{Y}^{3}\right)  \left\vert \text{\textquotedblleft}%
fail\text{\textquotedblright}\right\rangle _{Z}\\
+\left(  \left\vert OK\right\rangle _{Y}^{3}-\left\vert fail\right\rangle
_{Y}^{3}\right)  \left\vert \text{\textquotedblleft}OK\text{\textquotedblright%
}\right\rangle _{Z}%
\end{array}
\right\} \\
+\left\vert tails\right\rangle _{X}^{3}\left\{
\begin{array}
[c]{c}%
\left(  3\left\vert fail\right\rangle _{Y}^{3}+\left\vert OK\right\rangle
_{Y}^{3}\right)  \left\vert \text{\textquotedblleft}%
fail\text{\textquotedblright}\right\rangle _{Z}\\
-\left(  \left\vert OK\right\rangle _{Y}^{3}-\left\vert fail\right\rangle
_{Y}^{3}\right)  \left\vert \text{\textquotedblleft}OK\text{\textquotedblright%
}\right\rangle _{Z}%
\end{array}
\right\}
\end{array}
\right]
\end{multline}
What can Wigner legitimately infer about Xena's outcome at $t=0$? Prior to
$t=4$ he has yet to perform his own measurement of $w$ on $Y$, and prior to
$t=5$ he remains unaware of the outcome of Zeus's measurement of $z$ on $X$ at
$t=3$. But even before $t=4$\ Wigner can still use $\left\vert \psi
\right\rangle _{XYZ}^{3}$ to reason hypothetically about Xena's outcome,
conditional on Zeus's and his own measurements both having the outcome
\textquotedblleft OK\textquotedblright: on learning at $t=5$\ that these
antecedents are true, he can then infer the truth of the consequent of this
conditional. Wigner should therefore replace the incorrect reasoning of step
4* as follows.

\textit{Step 4} Assume Zeus's measurement of $z$ on $X$ at $t=3$ had a unique,
physical outcome and that there are then no interactions among $X$, $Y$, $Z$
prior to $t=4$. Then the state of $XYZ$ \textit{at} $t=4$ is%
\begin{equation}
\left\vert \psi\right\rangle _{XYZ}^{4}=\frac{1}{\sqrt{24}}\left[
\begin{array}
[c]{c}%
\left\vert heads\right\rangle _{X}^{4}\left\{
\begin{array}
[c]{c}%
\left(  3\left\vert fail\right\rangle _{Y}^{4}+\left\vert OK\right\rangle
_{Y}^{4}\right)  \left\vert \text{\textquotedblleft}%
fail\text{\textquotedblright}\right\rangle _{Z}^{4}\\
+\left(  \left\vert OK\right\rangle _{Y}^{4}-\left\vert fail\right\rangle
_{Y}^{4}\right)  \left\vert \text{\textquotedblleft}OK\text{\textquotedblright%
}\right\rangle _{Z}^{4}%
\end{array}
\right\}  \\
+\left\vert tails\right\rangle _{X}^{4}\left\{
\begin{array}
[c]{c}%
\left(  3\left\vert fail\right\rangle _{Y}^{4}+\left\vert OK\right\rangle
_{Y}^{4}\right)  \left\vert \text{\textquotedblleft}%
fail\text{\textquotedblright}\right\rangle _{Z}^{4}\\
-\left(  \left\vert OK\right\rangle _{Y}^{4}-\left\vert fail\right\rangle
_{Y}^{4}\right)  \left\vert \text{\textquotedblleft}OK\text{\textquotedblright%
}\right\rangle _{Z}^{4}%
\end{array}
\right\}
\end{array}
\right]  \label{XYZ state at t=4}%
\end{equation}
Suppose that Wigner's unique physical outcome on measuring $Y$ at $t=4$ were
\textquotedblleft OK\textquotedblright. Now consider the hypothesis that
Xena's outcome at $t=0$\ was \textquotedblleft tails\textquotedblright.
Equation (\ref{XYZ state at t=4}) then implies that the probability of
Wigner's outcome \textquotedblleft OK\textquotedblright\ would have been
$1/6$. On \ the alternative hypothesis that Xena's outcome at $t=0$\ was
\textquotedblleft heads\textquotedblright, equation (\ref{XYZ state at t=4})
also implies that the probability of Wigner getting outcome \textquotedblleft
OK\textquotedblright\ would have been $1/6$. So if Wigner were to get outcome
\textquotedblleft OK\textquotedblright\ for his measurement at $t=4$ his
knowledge of this outcome would not entitle him to infer the outcome of Xena's
measurement at $t=0$. Indeed, application of Bayes's theorem would lead him to
conclude that knowledge of the outcome of his measurement at $t=4$ should have
no effect on his estimate of the probabilities of Xena's possible outcomes:
they remain $prob(\text{\textquotedblleft}$heads$\text{\textquotedblright%
})=1/3,$ $prob($\textquotedblleft tails\textquotedblright$)=2/3$ after
conditionalizing on either possible outcome of his measurement at $t=4$, and
again after further conditionalizing on either possible outcome of Zeus's
measurement at $t=3$.

Consider, for purposes of contrast, how Wigner should reason if he knew that
Zeus performed no measurement at $t=3$. In that case he should assign the
following state to $XY$ at $t=4$:%
\begin{equation}
\left\vert \psi\right\rangle _{XY}^{4}=\frac{1}{\sqrt{3}}(\left\vert
heads\right\rangle _{X}^{4}\left\vert -1/2\right\rangle _{Y}^{4}+\sqrt
{2}\left\vert tails\right\rangle _{X}^{4}\left\vert fail\right\rangle _{Y}%
^{4}).
\end{equation}
Knowledge of the outcome \textquotedblleft OK\textquotedblright\ of his own
measurement of $w$ at $t=4$ would then entitle him to conclude (with
probability 1) that the outcome of Xena's measurement of $f$ on $c$ at $t=0$
was \textquotedblleft heads\textquotedblright. This conclusion follows by an
inference that parallels step 4* of the reasoning discussed previously. Unlike
step 4* itself, the parallel inference is valid because of the assumption that
Zeus performed no intervening measurement.

But that same assumption invalidates the premise of step 1 of the reasoning
discussed previously. Failing the conclusion of step 1, Wigner would no longer
be entitled to take steps 2 and 3. So if he knew that Zeus performed no
measurement at $t=3$ then Wigner could no longer validly conclude that the
outcome of Xena's measurement of $f$ on $c$ at $t=0$ was \textquotedblleft
tails\textquotedblright. In this contrasting case, Wigner should correctly,
and consistently, conclude that the unique outcome of Xena's measurement of
$f$ on $c$ at $t=0$ was \textquotedblleft heads\textquotedblright.

The preceding analysis of the two contrasting cases (with, and without, Zeus's
intervening measurement) shows clearly why \textit{Intervention Insensitivity}
must be rejected, as inconsistent with \textit{Universality} and \textit{No
collapse}. But it may appear to raise a worry about locality. For how can a
physically isolated intervening event like Zeus's distant measurement on $X$
have such an impact on Wigner's reasoning about local matters in this scenario?

The form of the question suggests an answer to the worry it seeks to express.
Zeus's measurement on $X$ certainly does not influence the outcome of Xena's
measurement: if it did, the influence would not be non-local but
time-reversed, since Zeus's measurement occurred later than Xena's! Xena's
outcome is what it is, irrespective of Zeus's measurement. If Zeus's
measurement were to influence anything it would be Wigner's outcome, not
Xena's. But Wigner's outcome \textquotedblleft OK\textquotedblright\ has the
same probability ($1/6$) whether or not Zeus performs his measurement. It is
only the correlation between Xena's and Wigner's outcomes that differs between
the cases where Zeus does and does not measure $z$ on $X$.

\qquad While Zeus's measurement modifies this correlation, it does so despite
being causally unrelated to any of its constituent events. This intervention
sensitivity is not an instance of non-local causal influence. The suspicion
that it is may arise from the view that correlations in non-separable states
like $\left\vert \psi\right\rangle _{XY}^{4}$ are causal because they specify
probabilistic counterfactual dependence between the outcomes of distant
measurements in violation of Bell inequalities \cite{Bell}. While controversy
continues \cite{Maudlin} as to whether such counterfactual dependence
constitutes or evidences non-local influence, there are well-known strategies
for denying that it does.\footnote{My preferred strategy \cite{Healey 2}
depends on an interventionist approach to causal influence.} So the failure of
\textit{Intervention Insensitivity} raises no \textit{new}\ worry about non-locality.

\section{A Third Argument}

I first heard this argument in a talk by Matthew Pusey \cite{Pusey}, who there
credits it to Luis Masanes. They should not be held responsible for my own
restatement and further development of the argument.

Once again, the argument is set in the context of a
\textit{Gedankenexperiment} featuring four experimenters. For variety I have
changed their names to Alice, Bob, Carol and Dan. But while Carol and Dan
perform difficult but technically feasible lab experiments, Alice and Bob are
credited with even more extreme abilities than the Zeus and Wigner who figured
in the previous arguments (though their exercise of these powers involves no
violation of unitary quantum theory).

Each of Alice and Bob are in their own separate laboratories, totally
physically isolated except for a shared Bohm-EPR pair of spin-$1/2$ particles
on which they intend to perform measurements of (normalized) spin-components,
one on each particle from the pair. Alice is to measure magnitude $A_{a}$
corresponding to operator $\hat{A}_{a}$ with eigenvalues $\{+1,-1\}$ on
particle $1$, while Bob is to measure magnitude $B_{b}$ corresponding to
operator $\hat{B}_{b}$ with eigenvalues $\{+1,-1\}$ on particle $2$. $a,b$
label two directions in space along which Alice and Bob (respectively) set the
axes of their spin-measuring devices. Alice will choose setting $a$ and
perform measurement of $A_{a}$ at spacelike-separation from Bob's choice of
setting $b$ and measurement of $B_{b}$.

But before performing these measurements, Alice and Bob first delegate a
similar task to their friends, Carol \ and Dan respectively. Carol occupies
her own separate laboratory, initially totally physically isolated from
Alice's: Dan occupies his own separate laboratory, initially totally
physically isolated from Bob's. Carol and Dan perform measurements on the
Bohm-EPR pair: Carol measures (normalized) spin-component $C_{c}$ on particle
$1$, while Dan measures (normalized) spin-component $D_{d}$ on particle $2$.
Carol's choice of setting $c$ and measurement of $C_{c}$ are each
spacelike-separated from Dan's choice of setting $d$ and measurement of
$D_{d}$. Assume Carol's and Dan's measurements each have a definite, physical
outcome that is registered, recorded and experienced by them separately in
their labs.

It is important to note that Alice and Carol both perform their measurements
on the very same particle $1$, and that Bob and Dan perform their measurements
on the very same particle $2$. To make this possible, after performing Carol's
measurement particle $1$ is transferred out of her lab and into Alice's lab,
and after performing Dan's measurement particle $2$ is transferred out of his
lab and into Bob's lab. Assume that measurement causes no physical
\textquotedblleft collapse\textquotedblright\ of the quantum state, so that
each spin-measurement proceeds in accordance with a unitary interaction
between the measured particle and the rest of the experimenter's lab, and that
this is consistent with its having a definite, physical outcome recorded by
the experimenter in that lab. It follows that Carol's measurement entangles
the state of her lab $C$ with that of particle $1$, while Dan's measurement
entangles the state of his lab $D$ with that of particle $2$.

But Alice and Bob use their superpowers to undo these entanglements by
applying very carefully tailored interactions, in the first case between $1$
and $C$, and in the second case\ between $2$ and $D$. This restores $C$ and
$D$ to their original states, and also restores the original spin-entangled
state of $1+2$. That is how it is possible for Alice and Bob to perform
spin-measurements on the same Bohm-EPR pair as Carol and Dan.

By assumption, we now have a situation in which successive measurements of
spin-component (in the $c$ and $a$ directions) have been performed on particle
$1$ of an individual Bohm-EPR pair, while successive measurements of
spin-component (in the $d$ and $b$ directions) have been performed on particle
$2$ of that pair. By assumption, each of these measurements has a definite,
physical outcome registered, recorded and experienced by an experimenter in
his or her laboratory. Finally suppose that this entire situation is repeated
very many times, each time with a different Bohm-EPR pair, giving rise to a
statistical distribution of results for the four outcomes in each trial.

We may use quantum theory to predict the corresponding probability
distribution by applying the Born rule to appropriate quantum states. From
Alice's perspective, events in a given trial unfold in the following sequence.
At time $t_{0}$ the particles are in state%
\begin{equation}
\left\vert \psi\right\rangle =\frac{1}{\sqrt{2}}(\left\vert \uparrow
\right\rangle _{1}\left\vert \downarrow\right\rangle _{2}-\left\vert
\downarrow\right\rangle _{1}\left\vert \uparrow\right\rangle _{2})
\end{equation}
while $C,D$ are in states $\left\vert ready\right\rangle _{C}$ , $\left\vert
ready\right\rangle _{D}$ respectively. Then Dan measures the $d$-spin of $2$
by means of a unitary interaction $\hat{U}_{D}^{2}$ as follows%
\begin{align}
\hat{U}_{D}^{2}\left\vert \downarrow_{d}\right\rangle _{2}\left\vert
\text{ready}\right\rangle _{D} &  =\left\vert d\text{-down}\right\rangle
_{2D}\\
\hat{U}_{D}^{2}\left\vert \uparrow_{d}\right\rangle _{2}\left\vert
\text{ready}\right\rangle _{D} &  =\left\vert d\text{-up}\right\rangle
_{2D}.\nonumber
\end{align}
So at time $t_{1}$ when Dan has recorded the definite outcome as either
$d$-down or $d$-up, Alice assigns the following state to $12D$%
\begin{equation}
\Psi_{1}=\frac{1}{\sqrt{2}}(\left\vert \uparrow_{d}\right\rangle
_{1}\left\vert d\text{-down}\right\rangle _{2D}-\left\vert \downarrow
_{d}\right\rangle _{1}\left\vert d\text{-up}\right\rangle _{2D}).
\end{equation}
Shortly after t$_{1}$ Carol measures the $c$-spin of particle $1$ by a unitary
interaction $\hat{U}_{C}^{1}$%
\begin{align}
\hat{U}_{C}^{1}\left\vert \downarrow_{c}\right\rangle _{1}\left\vert
\text{ready}\right\rangle _{C} &  =\left\vert c\text{-down}\right\rangle
_{1C}\\
\hat{U}_{C}^{1}\left\vert \uparrow_{c}\right\rangle _{1}\left\vert
\text{ready}\right\rangle _{C} &  =\left\vert c\text{-up}\right\rangle
_{1C}.\nonumber
\end{align}
So at time $t_{2}$ when Carol has recorded the definite outcome as either
$c$-down or $c$-up, Alice assigns the following state to $12DC$%
\begin{equation}
\Psi_{2}=\frac{1}{\sqrt{2}}(\hat{U}_{C}^{1}\left\vert \uparrow_{d}%
\right\rangle _{1}\left\vert \text{ready}\right\rangle _{C}\left\vert
d\text{-down}\right\rangle _{2D}-\hat{U}_{C}^{1}\left\vert \downarrow
_{d}\right\rangle _{1}\left\vert \text{ready}\right\rangle _{C}\left\vert
d\text{-up}\right\rangle _{2D})
\end{equation}
Next Alice \textquotedblleft undoes\textquotedblright\ the effects of Carol's
measurement by applying an interaction between $1$ and $C$ with unitary
$\hat{U}_{C}^{1\dag}$, and assigns the state $\Psi_{3}$\ at time $t_{3}$ to
$12D$ (which is no longer entangled with that of $C$)%
\begin{equation}
\Psi_{3}=\frac{1}{\sqrt{2}}(\left\vert \uparrow_{d}\right\rangle
_{1}\left\vert d\text{-down}\right\rangle _{2D}-\left\vert \downarrow
_{d}\right\rangle _{1}\left\vert d\text{-up}\right\rangle _{2D}).
\end{equation}
Shortly after $t_{3}$, Alice measures $a$-spin on $1$ and at time $t_{4\text{
}}$gets a definite, physical outcome of either $a$-down or $a$-up. Then Bob
\textquotedblleft undoes\textquotedblright\ Dan's measurement on particle $2$
by implementing an interaction in accordance with unitary $\hat{U}_{D}^{2\dag
}$, before measuring the $b$-spin of $2$ and at time $t_{5\text{ }}$getting a
definite, physical outcome of either $b$-down or $b$-up.

By applying the Born rule to the Bohm-EPR spin-state at $t_{0}$, Alice
predicts the probabilistic correlation function $E(c,d)$ for Carol's and Dan's
measurement outcomes%
\begin{equation}
E(c,d)=-\cos(c-d).
\end{equation}
To predict the correlation function $E(a,d)$ for Alice's and Dan's measurement
outcomes, Alice reasons as follows. If Carol had performed no measurement and
$C$ and $1$ had never interacted, then between $t_{1}$\ and $t_{4}$\ Alice and
Dan would just have been recording a correlation between outcomes of an
$a$-spin measurement on $1$ and an earlier $d$-spin measurement on $2$---a
standard Bell experiment with settings and measurements performed at timelike
separation. For such\ a case, the Born rule predicts%
\begin{equation}
E(a,d)=-\cos(a-d).
\end{equation}
In the present case, $C$ and $1$ interacted twice between $t_{1}$\ and $t_{4}
$, but these interactions had no overall effect on the state of the joint
system $12D$ at the time when Alice performed her measurement of $a$-spin: its
state was the same at $t_{3}$ as it had been at $t_{1}$ ($\Psi_{3}=\Psi_{1}$).
It follows that in the present case also quantum theory predicts the
correlation function%
\begin{equation}
E(a,d)=-\cos(a-d).
\end{equation}

After the effects of Dan's measurement on $2$ have been \textquotedblleft
undone\textquotedblright\ by Bob's implementation of the interaction $\hat
{U}_{D}^{2\dag}$, Alice should again recognize that the outcomes of her
measurement of $a$-spin on $1$ and Bob's spacelike-separated measurement of
$b$-spin on $2$ constitute a record of a correlation in a standard (spacelike
separated) Bell experiment, with predicted correlation function%
\begin{equation}
E(a,b)=-\cos(a-b).
\end{equation}

So far we have been considering the events involved in a single trial from
Alice's perspective. But those same events should also be considered from the
perspective of Bob. If the labs of Alice, Bob and friends are all in the same
state of motion, then the events we have considered will play out in the same
sequence also from Bob's perspective. But it is\ well known that the
\ time-order of spacelike separated events is not invariant under
transformations of inertial frame.

Suppose that Alice's lab and Carol's lab are in one state of motion relative
to frame $F$\ (moving to the right at speed $v$, say), while Bob's lab and
Carol's lab are in a different sate of motion (moving to the left at speed
$v$, say). To make sure that Alice is in position to manipulate $1$ and $C$ we
may assume that they both remain inside, and move together with, Alice's lab
$A$: and to make sure that Bob is in position to manipulate $2$ and $D$ we may
assume that they both remain inside, and move together with, Bob's lab $B$.
This arrangement is depicted in Figure 1. Relative to the state of motion of
Bob and Dan, the same events play out over a period marked by the sequence of
times $\left\langle t_{0}^{\ast},t_{1}^{\ast},t_{2}^{\ast},t_{3}^{\ast}%
,t_{4}^{\ast},t_{5}^{\ast}\right\rangle $.

Note that in the $^{\ast}$'d frame Carol's measurement precedes Dan's and
Bob's precedes Alice's. Most important, note that the state of $12C$ is the
same at $t_{3}^{\ast}$ as at $t_{1}^{\ast}$. Paralleling Alice's reasoning,
Bob should therefore conclude that in this situation the correlation function
for his outcome when measuring the $b$-spin of $2$ and Carol's outcome when
measuring the $c$-spin of $1$ is $E(b,c)=-\cos(b-c).$

It is a central assumption of this third argument that every spin measurement
by Alice, Bob, Carol or Dan has a definite, physical outcome---either spin up
or spin down with respect to the chosen direction. It follows that in a long
sequence of trials of the \textit{Gedankenexperiment} just described there
will be a statistical distribution of actual outcomes, with a set of outcomes
that may be labeled $\left\langle a,b,c,d\right\rangle $ in each trial.
Statistical correlations between pairs of actual experimental outcomes may be
represented in the usual way by statistical correlation functions $corr(a,b),$
$corr(b,c),$ $corr(c,d),$ $corr(a,d)$. It follows that these statistical
correlations will satisfy the inequality%
\begin{equation}
\left\vert corr(a,b)+corr(b,c)+corr(c,d)-corr(a,d)\right\vert \leq2.
\end{equation}
Note that no locality assumption is required to derive this inequality here,
since it is mathematically equivalent to the existence of a joint distribution
over the actual, physical outcomes whose existence has been assumed
\cite{Fine}.

But we saw that quantum mechanics predicts probabilistic correlation functions
$E(a,b),$ $E(b,c),$ $E(c,d),$ $E(a,d)$ for these pairs of outcomes that may be
compared to the inequality%
\begin{equation}
\left\vert E(a,b)+E(b,c)+E(c,d)-E(a,d)\right\vert \leq2. \label{BCHSH3}%
\end{equation}
It is well known that quantum theory predicts violation of inequality
(\ref{BCHSH3}) for certain choices of directions $a,b,c,d$. If the particles
and labs of the experimenters in the \textit{Gedankenexperiment} had been at
relative rest, then the choice of four directions in a plane defined by
rotations of $a%
{{}^\circ}%
=0%
{{}^\circ}%
,b%
{{}^\circ}%
=45%
{{}^\circ}%
,c%
{{}^\circ}%
=90%
{{}^\circ}%
,d%
{{}^\circ}%
=135%
{{}^\circ}%
$ from a fixed axis would yield maximal violation of (\ref{BCHSH3}) with
predicted value%
\begin{equation}
\left\vert E(a,b)+E(b,c)+E(c,d)-E(a,d)\right\vert =2\sqrt{2.}%
\end{equation}
The relativistic relative motion of labs and particles makes it necessary to
take account of the associated Wigner rotation of vectors, affecting the
predicted value for this choice of directions. But the inequality is still
maximally violated for a different choice of directions\footnote{\cite{Lee}
specifies the necessary directions in section 4. Rather than being coplanar
(with respect to frame $F$) these may be chosen to lie on a cone centered on
the direction of motion of the lab in which that spin measurement is
performed.}.\pagebreak

\section{Conclusion}

Each of the three arguments analyzed in this paper sought to establish a
contradiction between the universal applicability of unitary quantum theory
and the assumption that a well-conducted quantum measurement always has a
definite, physical outcome. The first argument succeeded in doing so only by
implicitly relying on assumptions that the work of Bell \cite{Bell} and Kochen
and Specker \cite{Kochen and Specker} gives us good reasons to reject---in
Einstein's \cite{Einstein} words, that in the circumstances described in the
associated \textit{Gedankenexperiment}\ the individual system (before the
measurement) has a definite value for all variables of the system, and more
specifically, \textit{that} value which is determined by a measurement of this
variable. Failing some such naive realist assumptions, nothing justifies the
argument's application of quantum theory to predict probabilities for outcomes
of hypothetical measurements which would be incompatible with those actually performed.

Though it does not rely on such naive realist assumptions, the second argument
also depends on a superficially plausible assumption about the outcomes of
counterfactual measurements\ I called \textit{intervention insensitivity},
according to which the truth-value of an outcome-counterfactual is insensitive
to the occurrence of a physically isolated intervening event. But in the
circumstances of the associated \textit{Gedankenexperiment}, intervention
insensitivity is itself incompatible with the universal applicability of
unitary quantum theory. Since a contradiction then follows even if each
(well-conducted) quantum measurement does \textit{not} have a definite,
physical outcome in the \textit{Gedankenexperiment}, the argument does not
establish its intended conclusion.

Unlike the first two arguments, the third argument relies on no implicit
assumptions about the outcomes of hypothetical measurements, since all the
outcomes it considers are assumed to be actual. I\ think it succeeds in
showing that, in the circumstances described in the associated
\textit{Gedankenexperiment}, the universal applicability of unitary quantum
theory implies (with probability approaching 1) that there is no consistent
assignment of values to the (supposedly definite, physical) outcomes of the
measurements in the sequence of trials there considered.

This result prompts further reflection on how to understand quantum theory.
But the circumstances of the \textit{Gedankenexperiment} in the third argument
are so extreme as forever to resist experimental realization. There are no
foreseeable circumstances in which the argument would require us to deny that
a well-conducted quantum measurement has a definite, physical outcome. The
arguments considered in this paper give us no reason to doubt the sincerity or
truth of experimenters' reports of definite, physical outcomes. But I\ think
the third argument should make us reconsider the extent and nature of their
objectivity. This paper was intended to both motivate and prepare the way for
the pursuit of that project. \newpage

\section{Appendix}

In restating the argument of (\cite{Renner}, \cite{Renner 2}) I have changed
the notation to try to make it easier to follow. The following table supplies
a translation between my notation and that used in \cite{Renner 2}.
\newline\newline%
\begin{tabular}
[c]{|c|c|c|c|c|}\hline
Agent & Lab & Measured System & Measured observable & Other observable\\\hline
$\bar{F}\leftrightsquigarrow$Xena & $\overline{L}\leftrightsquigarrow X$ &
$R\leftrightsquigarrow c$ & heads/tails$\leftrightsquigarrow f$ & \\\hline
$F\leftrightsquigarrow$Yvonne & $L\leftrightsquigarrow Y$ &
$S\leftrightsquigarrow s$ & up/down$\leftrightsquigarrow S_{z}$ & \\\hline
$\bar{W}\leftrightsquigarrow$Zeus & $\bullet\leftrightsquigarrow Z$ & $\bar
{L}\leftrightsquigarrow X$ & $\bar{w}\leftrightsquigarrow z$ & $\bullet
\leftrightsquigarrow x$\\\hline
$W\leftrightsquigarrow$Wigner & $\bullet\leftrightsquigarrow W$ &
$L\leftrightsquigarrow Y$ & $w\leftrightsquigarrow w$ & $\bullet
\leftrightsquigarrow y$\\\hline
\end{tabular}
\\[0.1in]

Readers familiar with Wigner's original \textquotedblleft
friend\textquotedblright\ argument \cite{Wigner} will be primed to attribute
extraordinary powers to the experimenter I have named Wigner, and I\ thought
it appropriate to name a second character with such almost \textquotedblleft
God-like\textquotedblright\ powers Zeus. This naturally suggested also giving
the experimenters charged with less extraordinary tasks names whose initial
letters are also at the end of the alphabet, with corresponding labels for
their labs and measured observables.

While such changes are merely cosmetic, my restatement deliberately lacks one
feature emphasized by the authors of the argument of \cite{Renner 2} that they
call \textquotedblleft consistent reasoning\textquotedblright, illustrate in
their Figure 1, and formalize in their assumption (C). Both in the original
and in my restatement it is Wigner ($W$) whose reasoning is the ultimate focus
of the argument. But the authors of the original argument consider it
important that Wigner's reasoning incorporates the reasoning of the other
experimenters (\textit{via} assumption (C)).

It is vital to check whether Wigner's reasoning is both internally consistent
and consistent with the reasoning of the other experimenters in this
\textit{Gedankenexperiment}. My restatement makes it clear how Wigner can
consistently apply quantum theory without considering the reasoning of any
other\ experimenters. But are the conclusions of this independent reasoning by
Wigner consistent with those of the other experimenters, based on their own
applications of quantum theory? Indeed they are, provided each experimenter
has applied quantum theory \textit{correctly}. The problem with the argument
of Frauchiger and Renner is that one experimenter (Xena/$\bar{F} $) has
applied quantum theory \textit{incorrectly}.

Recall step 4* of the reasoning in my restatement of this argument (see \S 3).
I attributed this reasoning to Wigner, while pointing out that Zeus's
subsequent measurement of $z$ renders it fallacious. Frauchiger and Renner
initially attribute parallel reasoning to Xena/$\bar{F}$ and then use
assumption (C) to attribute its conclusion also to $W$igner. To see where
things go wrong if Xena/$\bar{F}$ reasons this way, I quote from \cite{Renner
2}.\newpage

\begin{quote}
\textquotedblleft Specifically, agent $\bar{F}$ may start her reasoning with
the two statements%
\begin{align*}
s_{I}^{\bar{F}}  &  =\text{\textquotedblleft If }r=tails\text{ at time
}n:10\text{ then spin }S\text{ is in state }\left\vert \rightarrow
\right\rangle _{S}\text{ at time }n:10\text{\textquotedblright}\\
s_{M}^{\bar{F}}  &  =\text{\textquotedblleft The value }w\text{ is obtained by
a measurement of }L\text{ w.r.t.}\{\pi_{ok}^{H},\pi_{fail}^{H}%
\}\text{\textquotedblright.\textquotedblright}%
\end{align*}

\end{quote}

They conclude that $\bar{F}$ can infer from $s_{I}^{\bar{F}}$ and $s_{M}%
^{\bar{F}}$ that statement $s_{Q}^{\bar{F}}$ holds:%
\[
s_{Q}^{\bar{F}}=\text{ \textquotedblleft If }r=tails\text{ at time }n:10\text{
then I am certain that }W\text{ will observe }w=fail\text{ at }%
n:40\text{\textquotedblright.}%
\]

Starting with $s_{Q}^{\bar{F}}$, they then apply assumption (C) to the
reasoning of the other agents successively, eventually to establish that
$W$igner may conclude%
\[
s_{2}^{W}=\text{ \textquotedblleft If }\bar{w}=\overline{ok}\text{ at time
}n:30\text{ then I am certain that I will observe }w=fail\text{ at
}n:40\text{\textquotedblright,}%
\]
which (given (S)) is inconsistent with $W$'s independent conclusion (based on
assumption (Q))

\begin{quote}
$s_{Q}^{W}=$\textquotedblleft I am certain that there exists a round $n\in%
\mathbb{N}
_{\geq0}$ in which it is\newline\hspace*{0.4in}announced that $\bar
{w}=\overline{ok}$ at time $n:30$ and $w=ok$ at $n:40$.\textquotedblright
\end{quote}

But this chain of reasoning is based on a mistaken starting point, since
$\bar{F}$ has applied quantum theory incorrectly in asserting statement
$s_{Q}^{\bar{F}}$.\ Compare $s_{Q}^{\bar{F}}$ with the corresponding
conclusion of Wigner's fallacious reasoning in step 4* of \S 3:

\begin{quote}
\textquotedblleft If the unique outcome of Xena's measurement of $f$ on $c$ at
$t=0$ had been \textquotedblleft tails\textquotedblright, the unique outcome
of my measurement of $w$ on $Y$ at $t=4$ would have been \textquotedblleft
fail\textquotedblright.
\end{quote}

Agent $\bar{F}$'s reasoning was equally fallacious here. The problem starts
with statement $s_{I}^{\bar{F}}$:\ $\bar{F}$ is correct to assign state
$\left\vert \rightarrow\right\rangle _{S}$ to $S$\ at time $n:10$ for certain
purposes but not for others. Suppose, for example, that $\bar{F}$ had
\textquotedblleft flipped the quantum coin $R$\textquotedblright\ by passing
that system through the poles of a Stern-Gerlach magnet. By applying unitary
quantum theory, $\bar{F}$ should conclude that this will induce no physical
collapse of $R$'s spin state but entangle it with its translational state, and
thence with the rest of her lab \cite{Wigner Measurement}. So \ while $\bar
{F}$ would be correct then to assign state $\left\vert \rightarrow
\right\rangle _{S}$ to $S$\ at time $n:10$ for the purpose of predicting the
outcome of a \ subsequent spin measurement on $S$ alone, she would be
incorrect to assign state $\left\vert \rightarrow\right\rangle _{S}$ to
$S$\ at time $n:10$ for the purpose of predicting correlations between $S$ (or
anything with which it\ subsequently interacts) and her lab $\bar{L}$ (or
anything with which it\ subsequently interacts).

By using the phrase `is in', statement $s_{I}^{\bar{F}}$ ignores the essential
relativity of $S$'s\ state assignment at time $n:10$\ to these different
applications.\ By using $s_{I}^{\bar{F}}$ to infer $s_{M}^{\bar{F}} $, agent
$\bar{F}$ is, in effect, taking $\bar{F}$'s coin flip to involve the physical
collapse of $R$'s state rather than the unitary evolution represented by
equation (\ref{2}). So agent $\bar{F}$ is mistaken to assert $s_{Q}^{\bar{F}}%
$, and $W$ would be wrong to incorporate this mistake in his own reasoning by
applying assumption (C).

Frauchiger and Renner \cite{Renner 2} justify $\bar{F}$'s inference from
$s_{I}^{\bar{F}}$ and $s_{M}^{\bar{F}}$ to $s_{Q}^{\bar{F}}$\ by appeal to
assumption (Q). I have argued that $\bar{F}$ is not justified in asserting
$s_{Q}^{\bar{F}}$, since $\bar{F}$ is justified in using the state assignment
licensed by $s_{I}^{\bar{F}}$ for the purpose of predicting the outcome of a
measurement on $S$ only where $S$'s correlations with other systems (encoded
in an entangled state of a supersystem) may be neglected. But the sequence of
interactions in the \textit{Gedankenexperiment} successively entangle the
state of $S$ with those of $R$, $\bar{L}$, $L$ and $\bar{W}$. So in reasoning
about the outcome of $W$'s measurement of $w$, $\bar{F}$ must take account of
this progressive entanglement of the states of $S$ and $\bar{W}$.

Specifically, to predict the outcome of $W$'s measurement of $w$, $\bar{F}$
must represent that measurement as the second part of $W$'s joint measurement
on the system $\bar{W}+L$. This interaction between $W$ and $\bar{W}$\ was
represented in \S 3\ as the apparently innocuous Step 1 in which Wigner simply
asked Zeus what was the outcome of his measurement. But it is not this
interaction but the prior interaction between $\bar{W}$ and $L $ that
undercuts $\bar{F}$'s justification for using the state assignment $\left\vert
\rightarrow\right\rangle _{S}$ in inferrring $s_{Q}^{\bar{F}}$ from
$s_{I}^{\bar{F}}$ and $s_{M}^{\bar{F}}$. Only by neglecting the prior
interaction between $\bar{W}$ and $L$ can $\bar{F}$ draw the erroneous
conclusion $s_{Q}^{\bar{F}}$.

$W$igner can reason consistently about the unique, physical outcomes of all
experiments in the \textit{Gedankenexperiment} of (\cite{Renner}, \cite{Renner
2}) without any appeal to the reasoning of the other agents involved. Each of
these other agents may reason equally consistently. And their collective
reasoning is perfectly in accord with assumption (C) as well as the universal
applicability of unitary quantum theory and the existence of a unique,
physical outcome of every measurement that figures in the
\textit{Gedankenexperiment} of (\cite{Renner}, \cite{Renner 2}).

\begin{acknowledgement}
Thanks to Jeff Bub for a helpful correspondence on Frauchiger and Renner's
argument, to \v{C}aslav Brukner for conversations and correspondence over
several years, and to a reviewer for good strategic advice.\newpage
\end{acknowledgement}


\begin{thebibliography}{99}                                                                                               %


\bibitem {Fuchs}Fuchs, C. [2010] \textquotedblleft QBism, the perimeter of
quantum bayesianism\textquotedblright, arXiv:1003.5209.

\bibitem {FuchsMerminSchack}Fuchs, C., Mermin, N. D. and Schack, R. [2014]
\textquotedblleft An introduction to QBism with an application to the locality
of quantum mechanics\textquotedblright, \textit{American Journal of Physics}
\textbf{82}, 749--54.

\bibitem {Brukner}Brukner, \v{C}. [2017] \textquotedblleft On the quantum
measurement problem\textquotedblright, in Bertlmann, R. and Zeilinger, A.
(eds.) \textit{Quantum Unspeakables II}. Switzerland, Springer International, 95--117.

\bibitem {Brukner 2}Brukner, \v{C}. [2018] \textquotedblleft A No-Go Theorem
for Observer-Independent Facts\textquotedblright, \textit{Entropy}
\textbf{20}, 350.

\bibitem {Rovelli}Rovelli, C. [1996] \textquotedblleft Relational Quantum
Mechanics\textquotedblright, \textit{International Journal of Theoretical
Physics, }\textbf{35} 1637--78.

\bibitem {Popper}Popper, K.R. [1959] \textit{The Logic of Scientific
Discovery}. London: Hutchinson.

\bibitem {Deutsch}Deutsch, D. [1985] \textquotedblleft Quantum theory as a
universal physical theory\textquotedblright, \textit{International Journal of
Theoretical Physics} \textbf{24}, 1--41.

\bibitem {Wallace}Wallace, D. [2012] \textit{The Emergent Multiverse}. Oxford:
Oxford University Press.

\bibitem {Bell}Bell, J.S. [2004] \textit{Speakable and Unspeakable in Quantum
Mechanics}. Revised edition. Cambridge: Cambridge University Press.

\bibitem {Wigner}Wigner, E. [1961] \textquotedblleft Remarks on the Mind-body
question\textquotedblright, in Good, I. (ed.) \textit{The Scientist
Speculates}. London, Heinemann.

\bibitem {EPR}Einstein, A., Podolsky, B., and Rosen, N. [1935]
\textquotedblleft Can quantum-mechanical description of physical reality be
considered complete?\textquotedblright, \textit{Physical Review} \textbf{47}, 777--780.

\bibitem {Renner}Frauchiger, D. and Renner, R. [2016] "Single-world
interpretations of quantum theory cannot be self-consistent", https://arxiv.org/abs/1604.07422.

\bibitem {Renner 2}Frauchiger, D. and Renner, R. [forthcoming]
\textquotedblleft Quantum theory cannot consistently describe the use of
itself\textquotedblright.

\bibitem {Healey 2}Healey, R.A. [2016] "Locality, probability and causality",
in \textit{Quantum Nonlocality and Reality}, Mary Bell and Gao Shan, eds.
Cambridge: Cambridge University Press, 172--94.

\bibitem {Maudlin}Maudlin, T. [2011] \textit{Quantum Non-Locality and
Relativity}, 3rd edition. Chichester, West Sussex: Wiley-Blackwell.

\bibitem {Pusey}Pusey, M. [2016] \textquotedblleft Is QBism 80\% complete, or
20\%\textquotedblright, talk given at the \textit{Information-Theoretic
Interpretations of Quantum Mechanics} workshop, Western University, Canada
(available at
https://grfilms.net/v-matthew-pusey-is-qbism-80-complete-or-20-\_9Rs61l8MyY.html
).

\bibitem {Fine}Fine, A. [1982] \textquotedblleft Joint distributions, quantum
correlations, and commuting observables\textquotedblright. \textit{Journal of
Mathematical Physics} \textbf{23}, 1306--10.

\bibitem {Lee}Lee, D. and Chang-Young, E. [2004] \textquotedblleft Quantum
entanglement under Lorentz boost\textquotedblright, \textit{New Journal of
Physics }\textbf{6}, 67.

\bibitem {Kochen and Specker}Kochen, S. and Specker, E. [1967]
\textquotedblleft The problem of hidden variables in quantum
mechanics\textquotedblright, \textit{Journal of Mathematics and Mechanics}
\textbf{17}, 59--87.

\bibitem {Einstein}Einstein, A. [1949] \textquotedblleft Autobiographical
notes\textquotedblright\ in Schilpp, P. (ed.) \textit{Albert Einstein:
Philosopher-Scientist}. La Salle, IL: Open Court.

\bibitem {Wigner Measurement}Wigner, E. [1963] \textquotedblleft The problem
of measurement\textquotedblright, \textit{American Journal of Physics
}\textbf{31}, 6--15.
\end{thebibliography}
\end{document}